%% file: gol-decomp.tex
\documentclass[letterpaper]{article}

\usepackage{natbib,alifeconf}  %% The order is important
\usepackage[svgnames]{xcolor}
\usepackage[colorlinks,urlcolor=NavyBlue,citecolor=black!70,linkcolor=DarkSlateGrey]{hyperref}
\usepackage{amsmath}
\usepackage{amssymb}
\usepackage{mathtools}
\usepackage{url,hyperref,cleveref}
\usepackage{booktabs}

\usepackage[beginComment=\#]{algpseudocodex}

\newcommand{\ser}{\equiv} %state equivalence relation
\newcommand{\ter}{=} %transformation equivalence relation
\newcommand{\congclop}{{\mathbf C}} %congruence closure operator
\newcommand{\partop}{{\mathcal P}} %partition completion operator

\DeclareMathOperator{\image}{Im}

\DeclareMathOperator{\stab}{Stab}%^{\scalebox{.5}{$\subseteq$}}}%^\tinysubseteq}

\usepackage{todonotes}
\setuptodonotes{backgroundcolor=Chocolate!30,linecolor=black!30,inline}

\usepackage{amsthm}
\theoremstyle{plain}
\newtheorem{theorem}{Theorem}

\theoremstyle{definition}
\newtheorem{definition}[theorem]{Definition}

\usepackage{tikz}
\usetikzlibrary{automata, positioning, cd, shapes.geometric, backgrounds}
\tikzset{arr/.style={-Latex}}

\title{Measuring the Computational Power of Finite Patches of Cellular Automata}

\author{
  Attila Egri-Nagy$^{1}$, \and
  Chrystopher L. Nehaniv$^{2}$
    \mbox{}\\
    $^1$Akita International University, Japan \\
    $^2$University of Waterloo, Canada\\
    egri-nagy@aiu.ac.jp
} % email of corresponding author

\begin{document}

\maketitle

\begin{abstract}
  % Abstract length should not exceed 250 words
  Computational power can be measured by assigning an algebraic structure to a computational device. Here, we convert a small patch of Conway’s Game of Life into a transformation semigroup. The conversion captures not only time evolution but also interactive operations. In this way, the cellular automaton becomes directly programmable. 
  Once this measurement is made, we apply hierarchical decompositions to the resulting algebraic object as a way of understanding it. These decompositions are based on a macro/micro-state division inspired by statistical mechanics. However, cellular automata have a large number of global states. Therefore, we focus on partitioning the state space and creating morphic images—approximations that can serve as macro-level descriptions. The methods developed here are not limited to cellular automata; they apply more generally to discrete dynamical systems.
\end{abstract}

% If sharing code / data, anonymize your repository and paste the link here.
% Example of anonymizing sevice for github: https://anonymous.4open.science/
% delete this line if not needed
Data/Code available at: \url{https://codeberg.org/egri-nagy/gol-decomp}

%TOC just for now, to see the structure \tableofcontents

\section{Introduction}

We routinely compare the computational power of our devices.
For instance, we note that a laptop computer ten years ago had less memory and a slower processor than a new model today.
These comparisons are very informal and lack precision.
We might even draw the wrong conclusion (e.g., that there is no progress), as software tends to get bloated and thus it can hide hardware improvements.

Turning to the theory of computability, we have a sharp classification: a computing machine (defined by a programming language) can be Turing-complete or not.
Turing-completeness means having the same computational power as a Turing machine (\cite{sipser2013introduction}).
This condition turns out to be a low bar.
Once we have a conditional operation, we leave the realm of mere calculators and arrive at universal computers.
This is demonstrated by the numerous esoteric programming languages (\cite{temkin2025forty}), that are designed to be Turing-complete but utterly useless and so hard to use that it becomes entertaining.

Although ordinary programming models are often treated as Turing-complete in the abstract, any fixed physical device has finite resources and therefore realizes only a finite-state system.
No matter how big the memory is, we can always have an algorithmic problem and an input instance that will exceed that limit.
All physical computers are finite state machines, not Turing machines with infinite tapes. 
If we want to measure computational power better than a binary distinction, then we need a measure for finite computers.

By measurement we mean the assignment of a mathematical object to the thing we are measuring.
In simplest cases, this could be a number.
The size of the memory is a good proxy for measuring computational power, but here we aim for  more detailed objects, namely \emph{transformation semigroups}.
They are algebraic representations of finite state automata.
Transformation semigroups represent the totality of all the possible computations of an automaton.
Here, we will describe methods to measure the computational power of finite patches of cellular automata, and apply them to one of the most studied and well-known examples, Conway's Game of Life.

This is the first project in the research program described in \cite{compstructs2017}.
At that time there was no way to decompose large state set transformation semigroups.
This missing ingredient was a flexible decomposition method, which was finally constructed in  \cite{egrinagy2024relation}.

\subsection{Conway's Game of Life}

The Game of Life (\cite{izhikevich2015game}) is one of the most studied examples of cellular automata.
It is computationally universal \cite{wainright1974lifeuniversal}.
However, to build actual working computers, much engineering is needed (for details, see \cite{johnston2022conway}).
Here we want to harness the computational power of the cellular automaton directly, not by constructing a virtual machine on top of it.

\subsection{An algebraic view of state transition systems}

When modeling the global dynamics of a cellular automaton, we are interested in how events can be composed. 
In essence, we talk about \emph{semigroups}, sets closed under an associative
binary operation, which we may call composition.
However, we are interested in one particular representation,
as we want to model the global states of a finite patch.
The state-based representation is also the
one closest to automata theory.

\begin{definition}[Transformation Semigroup]
A \emph{transformation semigroup} $(X,S)$ is a finite nonempty set of
\emph{states} (points) $X$   and a set $S$ of total transformations of $X$, i.e.,
functions of type $X\rightarrow X$, closed under composition.
\end{definition}

We also say that $S$ \emph{acts} on $X$, expressed as a function of type
$X\times S \to X$.
For this action on the right we write $x\cdot s$, or simply $xs$.

The action is {\em faithful} if for all $s_1, s_2 \in S$, whenever \text{for all } $x \in X$, $x \cdot s_1 = x \cdot s_2$, then $s_1 = s_2$.
When the action is not faithful, we can identify transformations if they act the
same way on $X$, thus defining an equivalence relation $\equiv$.
Then the quotient semigroup $S/\!\!\equiv$ is faithful on $X$.

The \emph{image} of a transformation $s$ is $\image(s)=\{x\cdot s \mid x\in
X\}$.
The \emph{transporter} from set of states $A$ to another set of states $B$ is
the set of transformations $\{s\in S \mid A\cdot s\subseteq B\}$.
Elements of the transporter move the states of $A$ somewhere into $B$.
In general, it is a subset of $S$, not a subsemigroup.
We denote the transporter set made faithful by $S|^\ter_{A\rightarrow B}$.
The stabilizer of a set of states $A\subseteq X$ is the subsemigroup
$\stab_S(A)=\{s\in S\mid A\cdot s\subseteq A\}$, i.e., stabilizing $A$ in the
weaker sense of not leaving $A$.
%We write $\stab_S(x)$ instead of $\stab_S(\{x\})$.
We write $S|^\ter_A$ for $\stab_S(A)$ made faithful.
It is the image of a surjective homomorphism from a subsemigroup of $S$.
Such an image is often called
a \emph{divisor} of $S$.
A stabilizer is a special type of a transporter, one with identical source and
target.
Thus, $S|^\ter_{A}$ is a shorthand for $S\mid^\ter_{A\rightarrow A}$.

\section{Conversion}

Here we convert a finite patch of GoL to a transformation semigroup.
We use the global states of the patch to define the state set.
This approach, of course, limits the size of the patch.
For $n$ cells we have $2^n$ global states.

\subsection{Time evolution and interaction dynamics}
Interaction has long been argued to extend computation beyond the traditional algorithmic, input/output paradigm associated with Turing-machine computation, in the sense of going beyond the strong Church–Turing thesis \citep{Wegner1997WhyInteraction,GoldinSmolkaAttieSonderegger2004Turing,GoldinWegner2008InteractiveNature}. By contrast, repeated time evolution of a cellular automaton from a fixed initial state yields only the simplest finite transformation-semigroup dynamics: a transient followed by eventual periodic behaviour under the single generator given by the clock tick. Allowing external interventions, however, lets the system move between attractors and thereby generate much richer behaviour and greater algebraic complexity.

A cellular automaton patch started from a fixed configuration whose time evolution is due only by repeated clock ticks follows the familiar finite-state pattern of a transient leading into a periodic attractor cycle. Semigroup-theoretically, this is the simplest kind of transformation-semigroup dynamics, namely that of a monogenic semigroup generated by the single global update map. By adding interaction—here, operations such as setting or erasing cells—one can move between attractors rather than merely evolve within one basin, thereby substantially enriching the accessible dynamics and increasing the algebraic complexity of the generated transformation semigroup.

If we wish to use the patch as a computer, then it is not enough to consider its time evolution only.
Algebraically, the clock tick alone is just one generator, which we denote by $t$.
Cyclic (monogenic) transformation semigroups have an easy-to-describe structure: cycles with their corresponding basins of attractions.
This structure can be expressed in the attractor-cycle notation (\cite{egrinagy2024attractorcyclenotation}).

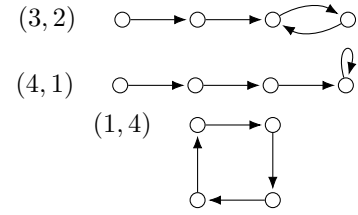
\begin{figure}
\begin{center}
  \begin{tikzpicture}[shorten >=1pt, node distance=1cm, on grid, auto,inner sep=2pt]
    \node[draw, circle] (s)   {};
    \node[draw, circle] (s2) [right of=s] {};
    \node[draw, circle] (s3) [right of=s2] {};
    \node[draw, circle] (s4) [right of=s3] {};
    \node(s5) [left of=s] {$(3,2)$};
    
    \path[->,every loop/.append style=-{Latex}]
    (s) edge [arr] (s2)
    (s2) edge [arr] (s3)
    (s3) edge [arr,bend left=30] (s4)
    (s4) edge [arr,bend left=30] (s3);

  \end{tikzpicture}

  \begin{tikzpicture}[shorten >=1pt, node distance=1cm, on grid, auto,inner sep=2pt]
    \node[draw, circle] (s)   {};
    \node[draw, circle] (s2) [right of=s] {};
    \node[draw, circle] (s3) [right of=s2] {};
    \node[draw, circle] (s4) [right of=s3] {};
    \node(s5) [left of=s] {$(4,1)$};
    
    \path[->,every loop/.append style=-{Latex}]
    (s) edge [arr] (s2)
    (s2) edge [arr] (s3)
    (s3) edge [arr] (s4)
    (s4) edge [arr,loop above] (s4);
  \end{tikzpicture}

    \begin{tikzpicture}[shorten >=1pt, node distance=1cm, on grid, auto,inner sep=2pt]
    \node[draw, circle] (s)   {};
    \node[draw, circle] (s2) [right of=s] {};
    \node[draw, circle] (s3) [below of=s2] {};
    \node[draw, circle] (s4) [left of=s3] {};
    \node(s5) [left of=s] {$(1,4)$};
    
    \path[->,every loop/.append style=-{Latex}]
    (s) edge [arr] (s2)
    (s2) edge [arr] (s3)
    (s3) edge [arr] (s4)
    (s4) edge [arr] (s);

  \end{tikzpicture}

\end{center}
  \caption{Schematic examples of semigroup elements with a given index-period pair.}
  \label{fig:indexperiod}
\end{figure}

The semigroup generated by a single element can be described by two numbers.
For an element $t$ in a finite semigroup $S$, we have $t^{m+r}=t^m$.
The \emph{index} of $t$ is
the smallest such integer $m\geq 1$ (the exponent of the first repeated power), and the \emph{period}  of $t$ is the smallest such integer $r\geq 1$ (the length of the cycle). For examples see Figure \ref{fig:indexperiod}.
We will use index-period analogously for an orbit of a state under the repeated action by a  single transformation.

Monogenic semigroups do not cover all possible dynamics.
This is an often overlooked mathematical fact: studying time evolution only is a very limited view of a system.
We need to interact with it in order to observe richer behaviour.

A standard use case of a cellular automaton GUI is drawing a configuration and then by switching on the updates we watch what evolves from the configuration.
Sometimes we stop to intervene by erasing or adding live cells.

Similar to graphical user interfaces we define the operations of setting and erasing cells.
Formally, setting the cell at position $(i,j)$ is denoted by $s_{ij}$, erasing by $e_{ij}$, and time evolution by $t$.  
Setting and erasing are both idempotent operations, i.e., repeating has the same effect as doing them once.
We allow arbitrary interleaving of these two types of operations.
A \emph{program} is simply a sequence of the above operations without any restriction, i.e., a sequence of $s_{ij}$'s, $e_{ij}$'s  and  $t$'s in any order.

\subsection{Spatial boundaries}

In theory, we can consider cellular automata virtually infinite.
In contrast, when working with small patches the boundaries and how we deal with them make decisive differences.
We consider finite boundaries, denoted by $\Box$, and toroidal boundaries, denoted by $\circlearrowright$.

One has to be careful with the toroidal for small patches.
The extreme case of $1\times 1$ demonstrates the possibility of the cell becoming the neighbor of itself.
For the $2\times 2$ patch the two methods give identical patches.

\subsection{$4\times 4$ patch}

A $4\times 4$ patch has 16 cells, thus the whole number of global states is $2^{16}=65536$.
For the toroidal patch, 47657 states are Garden-of-Eden states, the remaining 17879 states can come from other states (27.28\%).
For the finite patch, we have 11659 reachable states (17.79\%).

The time evolution transformation $t^\Box_{4\times 4}$ has index 14 and period 6.
The toroidal $t^\circlearrowright_{4\times 4}$ has index 9 and period 8, with a shorter transient but a longer cycle.
We look at the toroidal case in detail.
The biggest counter (cyclic group)  it has is $\mathbb{Z}_8$.
Consequently, it has $\mathbb{Z}_4$ and $\mathbb{Z}_2$.
There are no cycles with odd periods. Therefore, the biggest full transformation semigroups could be $T_2$.
In any case, looking at the cycles is interesting.
Period 1 configurations are still lifes, period 2 are blinkers.

\subsubsection{4-cycle: a `wide' glider}

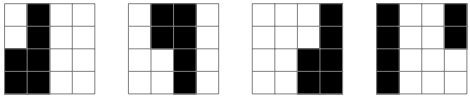
\begin{figure}
  \begin{center}
\begin{tikzpicture}[scale=0.3]
  \foreach \position in {(1,0), (1,1), (1,2), (1,3),(0,0),(0,1)}
    \draw[fill=black] \position rectangle +(1,1);
  \draw[step=1cm,very thin,gray] (0,0) grid (4,4);
\end{tikzpicture}
\hskip10pt
\begin{tikzpicture}[scale=0.3]
  \foreach \position in {(2,0), (2,1), (2,2), (2,3),(1,2),(1,3)}
    \draw[fill=black] \position rectangle +(1,1);
  \draw[step=1cm,very thin,gray] (0,0) grid (4,4);
\end{tikzpicture}
\hskip10pt
\begin{tikzpicture}[scale=0.3]
  \foreach \position in {(3,0), (3,1), (3,2), (3,3),(2,0),(2,1)}
    \draw[fill=black] \position rectangle +(1,1);
  \draw[step=1cm,very thin,gray] (0,0) grid (4,4);
  \end{tikzpicture}
  \hskip10pt
  \begin{tikzpicture}[scale=0.3]
  \foreach \position in {(0,0), (0,1), (0,2), (0,3),(3,2),(3,3)}
    \draw[fill=black] \position rectangle +(1,1);
  \draw[step=1cm,very thin,gray] (0,0) grid (4,4);
  \end{tikzpicture}
\end{center}
\caption{Wide glider on a $4\times 4$ toroidal patch. Time proceeds from left to right. This configuration yields a 4-cycle in the state space, a copy of $\mathbb{Z}_4$. The glider would continue on longer stripes of 4 cells.}
\label{fig:wide_glider}
\end{figure}

Looking at the 4-cycles we found a moving pattern, stretching between two opposite sides (see Figure \ref{fig:wide_glider}).
Giving it more room, we can see that it is using the toroidal coupling to swipe through a 4-cell-wide strip.

\subsubsection{8-cycles}

\begin{figure}
  \begin{center}
    \begin{tikzpicture}[scale=0.3]\foreach \position in {(2,3),(3,3),(3,1),(2,1),(3,2)}\draw[fill=black] \position rectangle +(1,1);\draw[step=1cm,very thin,gray] (0,0) grid (4,4);\end{tikzpicture}
\hskip10pt
\begin{tikzpicture}[scale=0.3]\foreach \position in {(2,3),(3,3),(0,2),(3,1),(2,1)}\draw[fill=black] \position rectangle +(1,1);\draw[step=1cm,very thin,gray] (0,0) grid (4,4);\end{tikzpicture}
\hskip10pt
\begin{tikzpicture}[scale=0.3]\foreach \position in {(3,3),(0,2),(3,1),(1,2)}\draw[fill=black] \position rectangle +(1,1);\draw[step=1cm,very thin,gray] (0,0) grid (4,4);\end{tikzpicture}
\hskip10pt
\begin{tikzpicture}[scale=0.3]\foreach \position in {(2,2),(0,3),(0,2),(3,2),(0,1)}\draw[fill=black] \position rectangle +(1,1);\draw[step=1cm,very thin,gray] (0,0) grid (4,4);\end{tikzpicture}
\vskip10pt
\begin{tikzpicture}[scale=0.3]\foreach \position in {(1,1),(1,3),(0,3),(0,2),(0,1)}\draw[fill=black] \position rectangle +(1,1);\draw[step=1cm,very thin,gray] (0,0) grid (4,4);\end{tikzpicture}
\hskip10pt
\begin{tikzpicture}[scale=0.3]\foreach \position in {(1,1),(1,3),(0,3),(3,2),(0,1)}\draw[fill=black] \position rectangle +(1,1);\draw[step=1cm,very thin,gray] (0,0) grid (4,4);\end{tikzpicture}
\hskip10pt
\begin{tikzpicture}[scale=0.3]\foreach \position in {(2,2),(0,3),(3,2),(0,1)}\draw[fill=black] \position rectangle +(1,1);\draw[step=1cm,very thin,gray] (0,0) grid (4,4);\end{tikzpicture}
\hskip10pt
\begin{tikzpicture}[scale=0.3]\foreach \position in {(3,3),(0,2),(3,1),(1,2),(3,2)}\draw[fill=black] \position rectangle +(1,1);\draw[step=1cm,very thin,gray] (0,0) grid (4,4);\end{tikzpicture}
\end{center}
\caption{A cycle of length 8 on the toroidal $4\times 4$ patch. The configurations are symmetric. }
\label{fig:8cycle}
\end{figure}
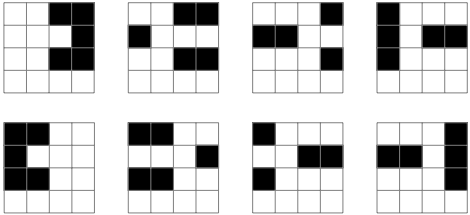

There are several 8-cycles. Figure \ref{fig:8cycle} shows one example with patterns that have a reflection symmetry.

\subsection{Growth due to Interaction}

We compute the size of the semigroups generated by the time evolution and the interactive transformations.
We do have advanced algorithms to compute the size from generators without fully enumerating the elements (\cite{EAST2019computing}).
However, depending on the structure of the semigroup, the algorithm may revert to complete enumeration.
Moreover, when a transformation consists of $2^{16}$ images, the required storage space is at least 256 kilobytes per semigroup element.
It is not feasible to compute the size of the semigroup of the $4\times 4$ patch when we include any of the interaction generators.
Therefore, we will use a $3\times 3$ patch to assess the growth of the number of transformations due to the addition of interactive transformations (Table \ref{tab:growth}).
\begin{table}
  \begin{center}
  \begin{tabular}{l|r}
    $S$ & $|S|$\\
    \hline
    $\langle t_{3\times 3}^{\circlearrowright}\rangle$ & 2\\
    $\langle t_{3\times 3}^{\circlearrowright},s_{00}\rangle$ & 11\\
    $\langle t_{3\times 3}^{\circlearrowright},s_{00},e_{00}\rangle$ & 35\\
    $\langle t_{3\times 3}^{\circlearrowright},s_{00},e_{00},s_{01},e_{01}\rangle$ & 551\\
    $\langle t_{3\times 3}^{\circlearrowright},s_{00},e_{00},s_{01},e_{01},s_{11},e_{11}\rangle$ & 18984\\
        $\langle s_{00},e_{00},s_{01},e_{01},s_{11},e_{11}\rangle$ & 26 
  \end{tabular}
\end{center}
\caption{Size of semigroups generated by time evolution and various interactive transformations. Only combining the two types leads to large semigroups.}
\label{tab:growth}
\end{table}
Using time evolution only yields a monogenic semigroup, which is small and has an elementary algebraic structure.
On the other hand, the interactive setting and erasing cell operations can create all configurations, but not through intrinsic computation.
This difference is decisive, but the distinction is not obvious.
An analogy can shed light on the difference: we can type into the computer that the 100th prime number is 541, or we can have a prime finder program outputting that.
In the second case, I can use the program to find other primes.
In algebra, the idempotents do not form cycles.

Our examples suggest that rich computational structure in this setting arises from the combination of intrinsic time evolution and interactive operations; neither component alone captures the full behaviour of interest here.
Therefore, we can draw one of our main conclusions:
\emph{time evolution and interactivity give computational power, and neither component suffices alone}.

\section{Simplified Views: Morphisms}

At this stage, we have the result of the measurement: the transformation semigroup assigned to the cellular automaton patch.
However, the semigroup is too big to investigate it directly.
We need to decompose it, and that involves creating simplified views.

A simplified view of a dynamical system is another one with less states and less transformations but with compatible dynamics.
This is an informal definition of  an algebraic morphism.
\subsection{Expressing Alikeness: Relational Morphisms}
For expressing relationships between transformation semigroups we use the usual idea
of structure preserving maps, the so-called homomorphisms.
However, we use relations instead of functions, and we need a pair of them for
the semigroup action.

\begin{definition}[Relational Morphism]
  A \emph{relational morphism} of transformation semigroups $(X,S)\xrightarrow{\theta,\varphi} (Y,T)$ is a pair of relations $(\theta: X\rightarrow Y, \varphi: S \rightarrow T)$ that are fully defined, i.e., $\theta(x)\neq\varnothing$ and $\varphi(s)\neq\varnothing$, and satisfy the condition of compatible actions for all $x\in X$ and $s\in S$:
$$y\in\theta(x), t\in \varphi(s) \implies y\cdot t \in \theta(x\cdot s),$$
or more succinctly: $\theta(x)\cdot\varphi(s)\subseteq \theta(x\cdot s)$, which
can be depicted by a subcommutative diagram.
\begin{center}
\begin{tikzcd}[row sep = large, column sep = large]
  X\times S \ar[d,"\theta\times \varphi"]
  \ar[r,"\cdot"]& X \ar[d,"\theta"] \ar[dl,phantom,"\subseteq"] \\
  Y\times T \ar[r, "\cdot"]& Y
\end{tikzcd}
\end{center}
\end{definition}

\subsection{Partitioning the state set}

We partition the large state set $X$, and use the semigroup
action on the classes $X/_\ser$ to get a
surjective morphism for the Covering Lemma method.
Arbitrary partitions do not work in general, as the semigroup action may not
respect the equivalence classes.
An equivalence relation $\ser$ on $X$ is a \emph{right congruence} if $x_1\ser
x_2\implies x_1s\ser x_2s$ for all $s\in S$, i.e., a transformation in the semigroup takes
all elements of an equivalence class to a single equivalence class.
%Therefore, the partition $X/_\ser$ defines a factor transformation semigroup,
%and by mapping onto this we get a surjective morphism
%$R_\ser:(X,S)\twoheadrightarrow S|^\ter_{X/_\ser}$.
%The image of the morphism is the factor semigroup made faithful.

Given a collection of disjoint subsets $\alpha=\{A_1,\ldots, A_k\}$,
we simply write $\bigcup\alpha$ as a shorthand for $\bigcup_{i=1}^kA_i$.
If this union is only a proper subset of $X$, $\bigcup \alpha \subset X$, then to have
a partition of $X$ we need to include the remaining elements as singleton sets.
This is a frequent operation, so we define this partition completion operator as
$\partop_X(\alpha)=\alpha\cup\{\{x\}\mid X\setminus\bigcup\alpha\}$.
To tell the equivalence class of an element we define $[x]=\{x'\mid x\ser x'\}$.
In computer algebra, we implement $[x]$ as lookup tables.

\subsubsection{Congruence from sets of identified states}
Constructing a congruence is not a trivial task.
We use a semi-automatic method:
we specify what states we want to identify,  and then use a closure algorithm to
find the finest partition (with the highest number of classes), in which the identified elements are equivalent.

First we define a function that decides whether a partition is a congruence or
not.
It systematically checks all classes, by applying the generators to all elements
in a class.
If two elements in the same class have images in different classes, it returns
those different classes.
As one step towards the congruence, they need to be merged into the same class.
If the input is a congruence already, it returns the empty set, meaning that
nothing needs to be merged.
\begin{algorithmic}
  \Function{ToBeMerged}{$\alpha,G$}
  
\Comment{ $\alpha$: partition of $X$,  \ \ $G$: generator set  for  $S$} 
\ForAll{$A \in \alpha$}
\ForAll{$g \in G$}
\State $M \leftarrow \{[x\cdot g] \mid x\in A\}$

\Comment{$M$ is the set of classes
  $g$ takes $A$ to}
\If{$|M|>1$}
\State \Return{$M$} \Comment{return the first split}
\EndIf
\EndFor
\EndFor
\State \Return{$\varnothing$} \Comment{$\alpha$ is a congruence, nothing to merge.}
\EndFunction
\end{algorithmic}
Merging is done by removing the classes to be merged from the partition, and adding
their union.
\begin{algorithmic}
\Function{Merge}{$\alpha,M$}\Comment{$\alpha$: partition of $X$, $M$: classes to be merged}
\State \Return{$(\alpha\setminus M)\cup(\bigcup M)$}
\EndFunction
\end{algorithmic}
We work gradually towards a congruence by repeatedly merging classes until there is nothing left to merge, thus the classes
are all compatible with the semigroup action.
\begin{algorithmic}
\Function{Congruence}{$I$, $G$}\Comment{$I$: disjoint subsets of $X$, $G$: generators for $S$}
\State $\alpha\leftarrow \partop_X(I)$\Comment{completing $I$ with singletons to form
a partition}
\Repeat
\State $M\leftarrow$ \Call{ToBeMerged}{$\alpha,G$}
\If{not $M=\varnothing$}
\State $\alpha\leftarrow$ \Call{Merge}{$\alpha,M$}
\EndIf
\Until{$M=\varnothing$}
\State \Return{$\alpha$}
\EndFunction
\end{algorithmic}
Note that this reference algorithm is written for simplicity of
presentation and the ease of understanding.
Some simple optimization tricks (e.g., randomizing the iterator $A\in\alpha$)
  may help if needed.
It is not optimized, since subsequent calls of \textsc{ToBeMerged} may check the same class again.

\subsubsection{Systematic Exploration of Congruences}
In practical problems, we may already have a good idea what states to identify.
Or, we can explore the results of random pairings systematically when we have a black box
system.

The set of all state congruences forms a partial order.
A partition $\alpha=\{A_1, \ldots, A_k\}$ is \emph{finer} than
$\beta=\{B_1,\ldots, B_l\}$, if each $A_i$ is a subset of some $B_j$, and we
write $\alpha\leq\beta$.
The top element is $X$ forming a single class.
The bottom is the set of all singletons $\{\{x\}\mid x\in X\}$.

The function \textsc{Congruence} above is a \emph{closure operator}.
Its domain is the set of partitions of the state set.
For the sake of brevity, we denote it by $\congclop$.
It satisfies the axioms of closure:
\begin{itemize}
\item \emph{extensive}, $\alpha\leq\congclop(\alpha)$;
\item \emph{increasing}, $\alpha\leq\beta \implies
    \congclop(\alpha)\leq\congclop(\beta)$;
\item\emph{idempotent}, $\congclop(\congclop(\alpha))=\congclop(\alpha).$
\end{itemize}

The partial order of the state congruences of $(X,S)$ forms a lattice.
The \emph{meet}, the greatest lower bound, can be
defined by intersection (and adding singleton classes for missing elements).
$$\alpha \wedge \beta= \partop_X(\alpha\cap\beta).$$
The \emph{join} $\vee$, the least upper bound, can be defined by the merging of those
classes that have non-empty intersections.
This requires a bit more description than simple union.
Let's define the relation between two sets of states as $A\sim B$ if $(A\cap
B)\neq\varnothing$.
This relation is reflexive, symmetric, but not transitive.
If we take the transitive closure, $\sim^+$, we get an equivalence relation.
We find the equivalence classes in $(\alpha\cup\beta)$ and take the unions of
sets in classes. This defines the join partition of $X$:

$$\alpha\vee\beta= \left\{\bigcup M\mid M\in (\alpha\cup\beta)/_{\sim^+}\right\}.$$

This is not a lattice where all the congruences are joins of atoms, since there
are examples with a single atom and several more congruences.
In $\langle 35344, 25512 \rangle$, the top is not a join of the atom.

\begin{figure*}
  \input{classOfGlider.tikz}

  \caption{The state congruence class of the 4 stages of the wide glider (98 configurations). In other words, these configurations form the smallest macro state that contains left-to-right glider. An open question is why precisely these states belong to the class.}
  \label{fig:class98}
\end{figure*}
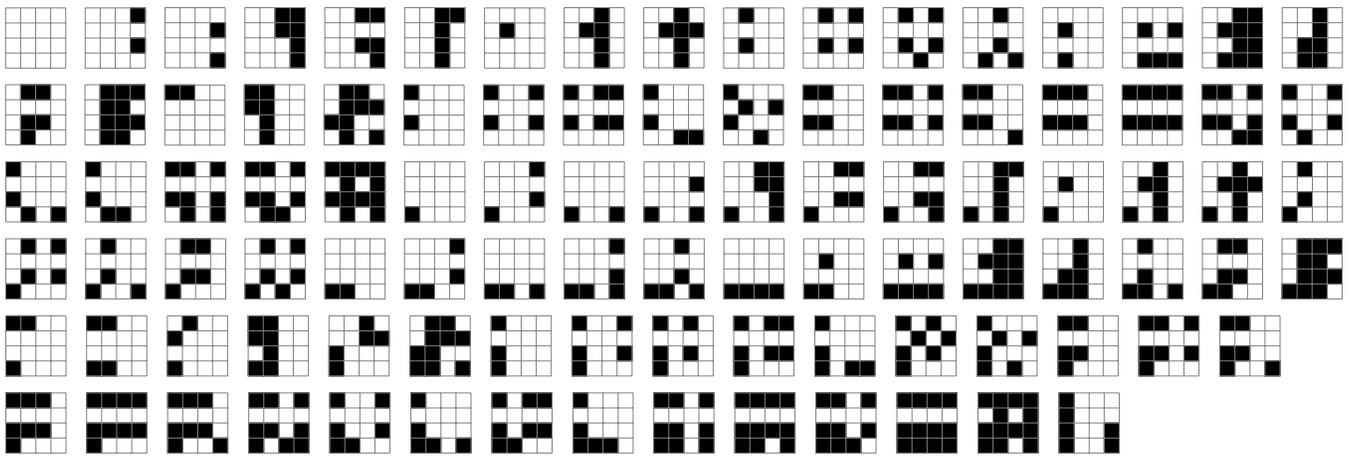

\subsubsection{Random Exploration}
We sampled randomly generated transformation semigroups with a given number of generators and degree.
We looked for `useful' state congruences in the sense that there are several congruence classes with more than one element.
The success of the congruence based method depends on
the size of the highest degree groups inside.
We see that with more generators, the semigroup is likely to have bigger degree
permutation groups, making this method inapplicable.
However, this observation is likely to be valid only for randomly generated transformation semigroups, since those do not explore the space of semigroups well.
For the Game of Life, only very few transformations relative to the complete space are allowed, thus finding congruences remains possible.

\subsection{The surjective morphism: $R_\ser(\theta_\ser,\varphi_\ser)$}
%Given a congruence $\ser$ on $X$, we define the transformation semigroup
%$(Y,T)$.
%First we give a surjective map on states $\theta_\ser:X\twoheadrightarrow Y$,
%then a surjective map on transformations $\varphi_\ser:S\twoheadrightarrow T$.

%Let's denote the partition corresponding to
%the equivalence classes $X/_\ser$ as $\alpha$.
First, we define the surjective map on states by
$\theta_\ser(x)=[x]$, i.e., the image of a state is the equivalence class
it belongs to.
%In computer algebra practice, these classes are recoded as positive integers,
%$P_i\rightarrow i$, and that is the state set we denote by $Y$.

%The semigroup $T$ comes from how $S$ acts on the partition $X/_\ser$.
Since $\ser$ is a congruence, all elements in a class go to the same class under
a transformation: $[x]\cdot\varphi_\ser(s)=[xs]$.
This defines the semigroup action $X/_\ser\times S\rightarrow X/_\ser$.
By making this action faithful we get $(X/_\ser,S|^=_{X/_\ser})$, which we
denote by $(Y,T)$.

The surjective map $(X,S)\rightarrow (Y,T)$ is the input for the flexible decomposition algorithm described in \cite{egrinagy2024relation}.

\subsection{Surjective Morphisms of the $4\times 4$ patch}

Since the semigroup is already too big, we need to decompose it.
We start with finding morphic images by searching for congruences.
What if we want to collapse the left-to-right wide glider (Figure \ref{fig:wide_glider}).
The state congruence class consists of 98 states (Figure \ref{fig:class98}).
We reduced the size of the state set from 65536 to 65439.
We can iterate this process.
Merging the states in an 8-cycle forms 2 more congruence classes, reducing the number of states to 65396.

We can also try to put all the Garden-of-Eden states into one macro-state.
In $\langle t_{4\times 4}^\circlearrowright, s_{00} \rangle$ there are 22,936 inaccessible states.
They indeed form a congruence class.
Another class with 42,424 microstates also appears.
The remaining 176 form singleton classes.
We now have a simplified view, a transformation semigroup on 178 states.

\section{Conclusions}

We described a method for measuring finite computational power precisely and applied it to cellular automata.
We also described a way to deal with large state sets algebraically.
Due to the unusual finiteness constraints we found some interesting new patterns in Conway's Game of Life.

Eventually, a mature mathematical theory will be able to tell the highest degree full transformation semigroup that can embed into an $m\times n$ patch.
This paper is the first step towards such a theory.

\section{Acknowledgements}

This project was funded in part by the Kakenhi grant
22K00015 by the Japan Society for the Promotion of Science (JSPS), titled `On progressing human understanding in the shadow of superhuman
deep learning artificial intelligence entities' (Grant-in-Aid for Scientific
Research type C, \url{https://kaken.nii.ac.jp/grant/KAKENHI-PROJECT-22K00015/}), and  in part by  the Natural Sciences and Engineering Research Council of Canada (NSERC), funding reference number RGPIN-2019-04669. 
%Cette recherche a \'et\'e financ\'ee en partie par le Conseil de recherches en sciences naturelles et en g\'enie du Canada (CRSNG), num\'ero de r\'ef\'erence RGPIN-2019-04669.  

\footnotesize
\bibliographystyle{apalike}
\bibliography{../coords} % replace by the name of your .bib file

\end{document}

%% file: classOfGlider.tikz
\begin{tikzpicture}[scale=0.2]
  \foreach \position in {}
    \draw[fill=black] \position rectangle +(1,1);
  \draw[step=1cm,very thin,gray] (0,0) grid (4,4);
  \end{tikzpicture}
\hskip5pt
\begin{tikzpicture}[scale=0.2]
  \foreach \position in {(3,3),(3,1)}
    \draw[fill=black] \position rectangle +(1,1);
  \draw[step=1cm,very thin,gray] (0,0) grid (4,4);
  \end{tikzpicture}
\hskip5pt
\begin{tikzpicture}[scale=0.2]
  \foreach \position in {(3,0),(3,2)}
    \draw[fill=black] \position rectangle +(1,1);
  \draw[step=1cm,very thin,gray] (0,0) grid (4,4);
  \end{tikzpicture}
\hskip5pt
\begin{tikzpicture}[scale=0.2]
  \foreach \position in {(2,2),(2,3),(3,3),(3,0),(3,1),(3,2)}
    \draw[fill=black] \position rectangle +(1,1);
  \draw[step=1cm,very thin,gray] (0,0) grid (4,4);
  \end{tikzpicture}
\hskip5pt
\begin{tikzpicture}[scale=0.2]
  \foreach \position in {(2,3),(3,3),(3,0),(3,1),(2,1)}
    \draw[fill=black] \position rectangle +(1,1);
  \draw[step=1cm,very thin,gray] (0,0) grid (4,4);
  \end{tikzpicture}
\hskip5pt
\begin{tikzpicture}[scale=0.2]
  \foreach \position in {(2,2),(2,3),(3,3),(2,0),(2,1)}
    \draw[fill=black] \position rectangle +(1,1);
  \draw[step=1cm,very thin,gray] (0,0) grid (4,4);
  \end{tikzpicture}
\hskip5pt
\begin{tikzpicture}[scale=0.2]
  \foreach \position in {(1,2)}
    \draw[fill=black] \position rectangle +(1,1);
  \draw[step=1cm,very thin,gray] (0,0) grid (4,4);
  \end{tikzpicture}
\hskip5pt
\begin{tikzpicture}[scale=0.2]
  \foreach \position in {(2,2),(2,3),(2,0),(2,1),(1,2)}
    \draw[fill=black] \position rectangle +(1,1);
  \draw[step=1cm,very thin,gray] (0,0) grid (4,4);
  \end{tikzpicture}
\hskip5pt
\begin{tikzpicture}[scale=0.2]
  \foreach \position in {(2,2),(2,3),(2,0),(2,1),(1,2),(3,2)}
    \draw[fill=black] \position rectangle +(1,1);
  \draw[step=1cm,very thin,gray] (0,0) grid (4,4);
  \end{tikzpicture}
\hskip5pt
\begin{tikzpicture}[scale=0.2]
  \foreach \position in {(1,1),(1,3)}
    \draw[fill=black] \position rectangle +(1,1);
  \draw[step=1cm,very thin,gray] (0,0) grid (4,4);
  \end{tikzpicture}
\hskip5pt
\begin{tikzpicture}[scale=0.2]
  \foreach \position in {(3,3),(1,1),(1,3),(3,1)}
    \draw[fill=black] \position rectangle +(1,1);
  \draw[step=1cm,very thin,gray] (0,0) grid (4,4);
  \end{tikzpicture}
\hskip5pt
\begin{tikzpicture}[scale=0.2]
  \foreach \position in {(3,3),(1,1),(1,3),(2,0),(3,1)}
    \draw[fill=black] \position rectangle +(1,1);
  \draw[step=1cm,very thin,gray] (0,0) grid (4,4);
  \end{tikzpicture}
\hskip5pt
\begin{tikzpicture}[scale=0.2]
  \foreach \position in {(1,0),(2,3),(3,0),(2,1)}
    \draw[fill=black] \position rectangle +(1,1);
  \draw[step=1cm,very thin,gray] (0,0) grid (4,4);
  \end{tikzpicture}
\hskip5pt
\begin{tikzpicture}[scale=0.2]
  \foreach \position in {(1,0),(1,2)}
    \draw[fill=black] \position rectangle +(1,1);
  \draw[step=1cm,very thin,gray] (0,0) grid (4,4);
  \end{tikzpicture}
\hskip5pt
\begin{tikzpicture}[scale=0.2]
  \foreach \position in {(1,0),(3,0),(2,0),(1,2),(3,2)}
    \draw[fill=black] \position rectangle +(1,1);
  \draw[step=1cm,very thin,gray] (0,0) grid (4,4);
  \end{tikzpicture}
\hskip5pt
\begin{tikzpicture}[scale=0.2]
  \foreach \position in {(2,2),(1,0),(2,3),(3,3),(3,0),(2,0),(3,1),(2,1),(1,2),(3,2)}
    \draw[fill=black] \position rectangle +(1,1);
  \draw[step=1cm,very thin,gray] (0,0) grid (4,4);
  \end{tikzpicture}
\hskip5pt
\begin{tikzpicture}[scale=0.2]
  \foreach \position in {(2,2),(1,0),(2,3),(1,1),(2,0),(2,1)}
    \draw[fill=black] \position rectangle +(1,1);
  \draw[step=1cm,very thin,gray] (0,0) grid (4,4);
  \end{tikzpicture}
\vskip5pt
\begin{tikzpicture}[scale=0.2]
  \foreach \position in {(1,0),(2,3),(1,1),(1,3),(2,1)}
    \draw[fill=black] \position rectangle +(1,1);
  \draw[step=1cm,very thin,gray] (0,0) grid (4,4);
  \end{tikzpicture}
\hskip5pt
\begin{tikzpicture}[scale=0.2]
  \foreach \position in {(2,2),(1,0),(2,3),(3,3),(1,1),(1,3),(2,0),(3,1),(2,1),(1,2)}
    \draw[fill=black] \position rectangle +(1,1);
  \draw[step=1cm,very thin,gray] (0,0) grid (4,4);
  \end{tikzpicture}
\hskip5pt
\begin{tikzpicture}[scale=0.2]
  \foreach \position in {(1,3),(0,3)}
    \draw[fill=black] \position rectangle +(1,1);
  \draw[step=1cm,very thin,gray] (0,0) grid (4,4);
  \end{tikzpicture}
\hskip5pt
\begin{tikzpicture}[scale=0.2]
  \foreach \position in {(1,0),(1,1),(1,3),(0,3),(0,2),(1,2)}
    \draw[fill=black] \position rectangle +(1,1);
  \draw[step=1cm,very thin,gray] (0,0) grid (4,4);
  \end{tikzpicture}
\hskip5pt
\begin{tikzpicture}[scale=0.2]
  \foreach \position in {(2,2),(1,0),(2,3),(1,1),(3,0),(1,3),(1,2),(3,2),(0,1)}
    \draw[fill=black] \position rectangle +(1,1);
  \draw[step=1cm,very thin,gray] (0,0) grid (4,4);
  \end{tikzpicture}
\hskip5pt
\begin{tikzpicture}[scale=0.2]
  \foreach \position in {(0,3),(0,1)}
    \draw[fill=black] \position rectangle +(1,1);
  \draw[step=1cm,very thin,gray] (0,0) grid (4,4);
  \end{tikzpicture}
\hskip5pt
\begin{tikzpicture}[scale=0.2]
  \foreach \position in {(3,3),(0,3),(3,1),(0,1)}
    \draw[fill=black] \position rectangle +(1,1);
  \draw[step=1cm,very thin,gray] (0,0) grid (4,4);
  \end{tikzpicture}
\hskip5pt
\begin{tikzpicture}[scale=0.2]
  \foreach \position in {(2,3),(3,3),(0,3),(3,1),(2,1),(0,1)}
    \draw[fill=black] \position rectangle +(1,1);
  \draw[step=1cm,very thin,gray] (0,0) grid (4,4);
  \end{tikzpicture}
\hskip5pt
\begin{tikzpicture}[scale=0.2]
  \foreach \position in {(3,0),(0,3),(2,0),(0,1)}
    \draw[fill=black] \position rectangle +(1,1);
  \draw[step=1cm,very thin,gray] (0,0) grid (4,4);
  \end{tikzpicture}
\hskip5pt
\begin{tikzpicture}[scale=0.2]
  \foreach \position in {(0,3),(2,0),(1,2),(3,2),(0,1)}
    \draw[fill=black] \position rectangle +(1,1);
  \draw[step=1cm,very thin,gray] (0,0) grid (4,4);
  \end{tikzpicture}
\hskip5pt
\begin{tikzpicture}[scale=0.2]
  \foreach \position in {(1,1),(1,3),(0,3),(0,1)}
    \draw[fill=black] \position rectangle +(1,1);
  \draw[step=1cm,very thin,gray] (0,0) grid (4,4);
  \end{tikzpicture}
\hskip5pt
\begin{tikzpicture}[scale=0.2]
  \foreach \position in {(3,3),(1,1),(1,3),(0,3),(3,1),(0,1)}
    \draw[fill=black] \position rectangle +(1,1);
  \draw[step=1cm,very thin,gray] (0,0) grid (4,4);
  \end{tikzpicture}
\hskip5pt
\begin{tikzpicture}[scale=0.2]
  \foreach \position in {(1,1),(3,0),(1,3),(0,3),(0,1)}
    \draw[fill=black] \position rectangle +(1,1);
  \draw[step=1cm,very thin,gray] (0,0) grid (4,4);
  \end{tikzpicture}
\hskip5pt
\begin{tikzpicture}[scale=0.2]
  \foreach \position in {(2,3),(1,1),(1,3),(0,3),(2,1),(0,1)}
    \draw[fill=black] \position rectangle +(1,1);
  \draw[step=1cm,very thin,gray] (0,0) grid (4,4);
  \end{tikzpicture}
\hskip5pt
\begin{tikzpicture}[scale=0.2]
  \foreach \position in {(2,3),(3,3),(1,1),(1,3),(0,3),(3,1),(2,1),(0,1)}
    \draw[fill=black] \position rectangle +(1,1);
  \draw[step=1cm,very thin,gray] (0,0) grid (4,4);
  \end{tikzpicture}
\hskip5pt
\begin{tikzpicture}[scale=0.2]
  \foreach \position in {(3,3),(1,1),(3,0),(1,3),(0,3),(2,0),(3,1),(0,1)}
    \draw[fill=black] \position rectangle +(1,1);
  \draw[step=1cm,very thin,gray] (0,0) grid (4,4);
  \end{tikzpicture}
\hskip5pt
\begin{tikzpicture}[scale=0.2]
  \foreach \position in {(1,0),(3,3),(0,3),(3,1),(0,1)}
    \draw[fill=black] \position rectangle +(1,1);
  \draw[step=1cm,very thin,gray] (0,0) grid (4,4);
  \end{tikzpicture}
\vskip5pt
\begin{tikzpicture}[scale=0.2]
  \foreach \position in {(1,0),(3,0),(0,3),(0,1)}
    \draw[fill=black] \position rectangle +(1,1);
  \draw[step=1cm,very thin,gray] (0,0) grid (4,4);
  \end{tikzpicture}
\hskip5pt
\begin{tikzpicture}[scale=0.2]
  \foreach \position in {(1,0),(0,3),(2,0),(0,1)}
    \draw[fill=black] \position rectangle +(1,1);
  \draw[step=1cm,very thin,gray] (0,0) grid (4,4);
  \end{tikzpicture}
\hskip5pt
\begin{tikzpicture}[scale=0.2]
  \foreach \position in {(1,0),(3,3),(1,1),(3,0),(1,3),(0,3),(3,1),(0,1)}
    \draw[fill=black] \position rectangle +(1,1);
  \draw[step=1cm,very thin,gray] (0,0) grid (4,4);
  \end{tikzpicture}
\hskip5pt
\begin{tikzpicture}[scale=0.2]
  \foreach \position in {(1,0),(3,3),(1,1),(1,3),(0,3),(2,0),(3,1),(0,1)}
    \draw[fill=black] \position rectangle +(1,1);
  \draw[step=1cm,very thin,gray] (0,0) grid (4,4);
  \end{tikzpicture}
\hskip5pt
\begin{tikzpicture}[scale=0.2]
  \foreach \position in {(1,0),(2,3),(3,3),(1,1),(3,0),(1,3),(0,3),(3,1),(2,1),(1,2),(3,2),(0,1)}
    \draw[fill=black] \position rectangle +(1,1);
  \draw[step=1cm,very thin,gray] (0,0) grid (4,4);
  \end{tikzpicture}
\hskip5pt
\begin{tikzpicture}[scale=0.2]
  \foreach \position in {(0,0)}
    \draw[fill=black] \position rectangle +(1,1);
  \draw[step=1cm,very thin,gray] (0,0) grid (4,4);
  \end{tikzpicture}
\hskip5pt
\begin{tikzpicture}[scale=0.2]
  \foreach \position in {(0,0),(3,3),(3,1)}
    \draw[fill=black] \position rectangle +(1,1);
  \draw[step=1cm,very thin,gray] (0,0) grid (4,4);
  \end{tikzpicture}
\hskip5pt
\begin{tikzpicture}[scale=0.2]
  \foreach \position in {(0,0),(3,0)}
    \draw[fill=black] \position rectangle +(1,1);
  \draw[step=1cm,very thin,gray] (0,0) grid (4,4);
  \end{tikzpicture}
\hskip5pt
\begin{tikzpicture}[scale=0.2]
  \foreach \position in {(0,0),(3,0),(3,2)}
    \draw[fill=black] \position rectangle +(1,1);
  \draw[step=1cm,very thin,gray] (0,0) grid (4,4);
  \end{tikzpicture}
\hskip5pt
\begin{tikzpicture}[scale=0.2]
  \foreach \position in {(2,2),(0,0),(2,3),(3,3),(3,0),(3,1),(3,2)}
    \draw[fill=black] \position rectangle +(1,1);
  \draw[step=1cm,very thin,gray] (0,0) grid (4,4);
  \end{tikzpicture}
\hskip5pt
\begin{tikzpicture}[scale=0.2]
  \foreach \position in {(0,0),(2,3),(3,3),(3,1),(2,1)}
    \draw[fill=black] \position rectangle +(1,1);
  \draw[step=1cm,very thin,gray] (0,0) grid (4,4);
  \end{tikzpicture}
\hskip5pt
\begin{tikzpicture}[scale=0.2]
  \foreach \position in {(0,0),(2,3),(3,3),(3,0),(3,1),(2,1)}
    \draw[fill=black] \position rectangle +(1,1);
  \draw[step=1cm,very thin,gray] (0,0) grid (4,4);
  \end{tikzpicture}
\hskip5pt
\begin{tikzpicture}[scale=0.2]
  \foreach \position in {(2,2),(0,0),(2,3),(3,3),(2,0),(2,1)}
    \draw[fill=black] \position rectangle +(1,1);
  \draw[step=1cm,very thin,gray] (0,0) grid (4,4);
  \end{tikzpicture}
\hskip5pt
\begin{tikzpicture}[scale=0.2]
  \foreach \position in {(0,0),(1,2)}
    \draw[fill=black] \position rectangle +(1,1);
  \draw[step=1cm,very thin,gray] (0,0) grid (4,4);
  \end{tikzpicture}
\hskip5pt
\begin{tikzpicture}[scale=0.2]
  \foreach \position in {(2,2),(0,0),(2,3),(2,0),(2,1),(1,2)}
    \draw[fill=black] \position rectangle +(1,1);
  \draw[step=1cm,very thin,gray] (0,0) grid (4,4);
  \end{tikzpicture}
\hskip5pt
\begin{tikzpicture}[scale=0.2]
  \foreach \position in {(2,2),(0,0),(2,3),(2,0),(2,1),(1,2),(3,2)}
    \draw[fill=black] \position rectangle +(1,1);
  \draw[step=1cm,very thin,gray] (0,0) grid (4,4);
  \end{tikzpicture}
\hskip5pt
\begin{tikzpicture}[scale=0.2]
  \foreach \position in {(0,0),(1,1),(1,3)}
    \draw[fill=black] \position rectangle +(1,1);
  \draw[step=1cm,very thin,gray] (0,0) grid (4,4);
  \end{tikzpicture}
\vskip5pt
\begin{tikzpicture}[scale=0.2]
  \foreach \position in {(0,0),(3,3),(1,1),(1,3),(3,1)}
    \draw[fill=black] \position rectangle +(1,1);
  \draw[step=1cm,very thin,gray] (0,0) grid (4,4);
  \end{tikzpicture}
\hskip5pt
\begin{tikzpicture}[scale=0.2]
  \foreach \position in {(0,0),(1,1),(3,0),(1,3)}
    \draw[fill=black] \position rectangle +(1,1);
  \draw[step=1cm,very thin,gray] (0,0) grid (4,4);
  \end{tikzpicture}
\hskip5pt
\begin{tikzpicture}[scale=0.2]
  \foreach \position in {(0,0),(2,3),(1,1),(1,3),(2,1)}
    \draw[fill=black] \position rectangle +(1,1);
  \draw[step=1cm,very thin,gray] (0,0) grid (4,4);
  \end{tikzpicture}
\hskip5pt
\begin{tikzpicture}[scale=0.2]
  \foreach \position in {(0,0),(3,3),(1,1),(1,3),(2,0),(3,1)}
    \draw[fill=black] \position rectangle +(1,1);
  \draw[step=1cm,very thin,gray] (0,0) grid (4,4);
  \end{tikzpicture}
\hskip5pt
\begin{tikzpicture}[scale=0.2]
  \foreach \position in {(0,0),(1,0)}
    \draw[fill=black] \position rectangle +(1,1);
  \draw[step=1cm,very thin,gray] (0,0) grid (4,4);
  \end{tikzpicture}
\hskip5pt
\begin{tikzpicture}[scale=0.2]
  \foreach \position in {(0,0),(1,0),(3,3),(3,1)}
    \draw[fill=black] \position rectangle +(1,1);
  \draw[step=1cm,very thin,gray] (0,0) grid (4,4);
  \end{tikzpicture}
\hskip5pt
\begin{tikzpicture}[scale=0.2]
  \foreach \position in {(0,0),(1,0),(3,0)}
    \draw[fill=black] \position rectangle +(1,1);
  \draw[step=1cm,very thin,gray] (0,0) grid (4,4);
  \end{tikzpicture}
\hskip5pt
\begin{tikzpicture}[scale=0.2]
  \foreach \position in {(0,0),(1,0),(3,3),(3,0),(3,1)}
    \draw[fill=black] \position rectangle +(1,1);
  \draw[step=1cm,very thin,gray] (0,0) grid (4,4);
  \end{tikzpicture}
\hskip5pt
\begin{tikzpicture}[scale=0.2]
  \foreach \position in {(0,0),(1,0),(2,3),(3,0),(2,1)}
    \draw[fill=black] \position rectangle +(1,1);
  \draw[step=1cm,very thin,gray] (0,0) grid (4,4);
  \end{tikzpicture}
\hskip5pt
\begin{tikzpicture}[scale=0.2]
  \foreach \position in {(0,0),(1,0),(3,0),(2,0)}
    \draw[fill=black] \position rectangle +(1,1);
  \draw[step=1cm,very thin,gray] (0,0) grid (4,4);
  \end{tikzpicture}
\hskip5pt
\begin{tikzpicture}[scale=0.2]
  \foreach \position in {(0,0),(1,0),(1,2)}
    \draw[fill=black] \position rectangle +(1,1);
  \draw[step=1cm,very thin,gray] (0,0) grid (4,4);
  \end{tikzpicture}
\hskip5pt
\begin{tikzpicture}[scale=0.2]
  \foreach \position in {(0,0),(1,0),(3,0),(2,0),(1,2),(3,2)}
    \draw[fill=black] \position rectangle +(1,1);
  \draw[step=1cm,very thin,gray] (0,0) grid (4,4);
  \end{tikzpicture}
\hskip5pt
\begin{tikzpicture}[scale=0.2]
  \foreach \position in {(2,2),(0,0),(1,0),(2,3),(3,3),(3,0),(2,0),(3,1),(2,1),(1,2),(3,2)}
    \draw[fill=black] \position rectangle +(1,1);
  \draw[step=1cm,very thin,gray] (0,0) grid (4,4);
  \end{tikzpicture}
\hskip5pt
\begin{tikzpicture}[scale=0.2]
  \foreach \position in {(2,2),(0,0),(1,0),(2,3),(1,1),(2,0),(2,1)}
    \draw[fill=black] \position rectangle +(1,1);
  \draw[step=1cm,very thin,gray] (0,0) grid (4,4);
  \end{tikzpicture}
\hskip5pt
\begin{tikzpicture}[scale=0.2]
  \foreach \position in {(0,0),(1,0),(1,1),(3,0),(1,3)}
    \draw[fill=black] \position rectangle +(1,1);
  \draw[step=1cm,very thin,gray] (0,0) grid (4,4);
  \end{tikzpicture}
\hskip5pt
\begin{tikzpicture}[scale=0.2]
  \foreach \position in {(0,0),(1,0),(2,3),(1,1),(1,3),(2,1)}
    \draw[fill=black] \position rectangle +(1,1);
  \draw[step=1cm,very thin,gray] (0,0) grid (4,4);
  \end{tikzpicture}
\hskip5pt
\begin{tikzpicture}[scale=0.2]
  \foreach \position in {(2,2),(0,0),(1,0),(2,3),(3,3),(1,1),(1,3),(2,0),(3,1),(2,1),(1,2)}
    \draw[fill=black] \position rectangle +(1,1);
  \draw[step=1cm,very thin,gray] (0,0) grid (4,4);
  \end{tikzpicture}
\vskip5pt
\begin{tikzpicture}[scale=0.2]
  \foreach \position in {(0,0),(1,3),(0,3)}
    \draw[fill=black] \position rectangle +(1,1);
  \draw[step=1cm,very thin,gray] (0,0) grid (4,4);
  \end{tikzpicture}
\hskip5pt
\begin{tikzpicture}[scale=0.2]
  \foreach \position in {(0,0),(1,0),(1,3),(0,3)}
    \draw[fill=black] \position rectangle +(1,1);
  \draw[step=1cm,very thin,gray] (0,0) grid (4,4);
  \end{tikzpicture}
\hskip5pt
\begin{tikzpicture}[scale=0.2]
  \foreach \position in {(0,0),(1,3),(0,2)}
    \draw[fill=black] \position rectangle +(1,1);
  \draw[step=1cm,very thin,gray] (0,0) grid (4,4);
  \end{tikzpicture}
\hskip5pt
\begin{tikzpicture}[scale=0.2]
  \foreach \position in {(0,0),(1,0),(1,1),(1,3),(0,3),(0,2),(1,2)}
    \draw[fill=black] \position rectangle +(1,1);
  \draw[step=1cm,very thin,gray] (0,0) grid (4,4);
  \end{tikzpicture}
\hskip5pt
\begin{tikzpicture}[scale=0.2]
  \foreach \position in {(2,2),(0,0),(2,3),(3,2),(0,1)}
    \draw[fill=black] \position rectangle +(1,1);
  \draw[step=1cm,very thin,gray] (0,0) grid (4,4);
  \end{tikzpicture}
\hskip5pt
\begin{tikzpicture}[scale=0.2]
  \foreach \position in {(2,2),(0,0),(1,0),(2,3),(1,1),(3,0),(1,3),(1,2),(3,2),(0,1)}
    \draw[fill=black] \position rectangle +(1,1);
  \draw[step=1cm,very thin,gray] (0,0) grid (4,4);
  \end{tikzpicture}
\hskip5pt
\begin{tikzpicture}[scale=0.2]
  \foreach \position in {(0,0),(0,3),(0,1)}
    \draw[fill=black] \position rectangle +(1,1);
  \draw[step=1cm,very thin,gray] (0,0) grid (4,4);
  \end{tikzpicture}
\hskip5pt
\begin{tikzpicture}[scale=0.2]
  \foreach \position in {(0,0),(3,3),(0,3),(3,1),(0,1)}
    \draw[fill=black] \position rectangle +(1,1);
  \draw[step=1cm,very thin,gray] (0,0) grid (4,4);
  \end{tikzpicture}
\hskip5pt
\begin{tikzpicture}[scale=0.2]
  \foreach \position in {(0,0),(2,3),(0,3),(2,1),(0,1)}
    \draw[fill=black] \position rectangle +(1,1);
  \draw[step=1cm,very thin,gray] (0,0) grid (4,4);
  \end{tikzpicture}
\hskip5pt
\begin{tikzpicture}[scale=0.2]
  \foreach \position in {(0,0),(2,3),(3,3),(0,3),(3,1),(2,1),(0,1)}
    \draw[fill=black] \position rectangle +(1,1);
  \draw[step=1cm,very thin,gray] (0,0) grid (4,4);
  \end{tikzpicture}
\hskip5pt
\begin{tikzpicture}[scale=0.2]
  \foreach \position in {(0,0),(3,0),(0,3),(2,0),(0,1)}
    \draw[fill=black] \position rectangle +(1,1);
  \draw[step=1cm,very thin,gray] (0,0) grid (4,4);
  \end{tikzpicture}
\hskip5pt
\begin{tikzpicture}[scale=0.2]
  \foreach \position in {(0,0),(2,3),(0,3),(2,1),(1,2),(3,2),(0,1)}
    \draw[fill=black] \position rectangle +(1,1);
  \draw[step=1cm,very thin,gray] (0,0) grid (4,4);
  \end{tikzpicture}
\hskip5pt
\begin{tikzpicture}[scale=0.2]
  \foreach \position in {(0,0),(0,3),(2,0),(1,2),(3,2),(0,1)}
    \draw[fill=black] \position rectangle +(1,1);
  \draw[step=1cm,very thin,gray] (0,0) grid (4,4);
  \end{tikzpicture}
\hskip5pt
\begin{tikzpicture}[scale=0.2]
  \foreach \position in {(0,0),(1,1),(1,3),(0,3),(0,1)}
    \draw[fill=black] \position rectangle +(1,1);
  \draw[step=1cm,very thin,gray] (0,0) grid (4,4);
  \end{tikzpicture}
\hskip5pt
\begin{tikzpicture}[scale=0.2]
  \foreach \position in {(0,0),(3,3),(1,1),(1,3),(0,3),(3,1),(0,1)}
    \draw[fill=black] \position rectangle +(1,1);
  \draw[step=1cm,very thin,gray] (0,0) grid (4,4);
  \end{tikzpicture}
\hskip5pt
\begin{tikzpicture}[scale=0.2]
  \foreach \position in {(0,0),(1,1),(3,0),(1,3),(0,3),(0,1)}
    \draw[fill=black] \position rectangle +(1,1);
  \draw[step=1cm,very thin,gray] (0,0) grid (4,4);
  \end{tikzpicture}
\vskip5pt
\begin{tikzpicture}[scale=0.2]
  \foreach \position in {(0,0),(2,3),(1,1),(1,3),(0,3),(2,1),(0,1)}
    \draw[fill=black] \position rectangle +(1,1);
  \draw[step=1cm,very thin,gray] (0,0) grid (4,4);
  \end{tikzpicture}
\hskip5pt
\begin{tikzpicture}[scale=0.2]
  \foreach \position in {(0,0),(2,3),(3,3),(1,1),(1,3),(0,3),(3,1),(2,1),(0,1)}
    \draw[fill=black] \position rectangle +(1,1);
  \draw[step=1cm,very thin,gray] (0,0) grid (4,4);
  \end{tikzpicture}
\hskip5pt
\begin{tikzpicture}[scale=0.2]
  \foreach \position in {(0,0),(2,3),(1,1),(3,0),(1,3),(0,3),(2,1),(0,1)}
    \draw[fill=black] \position rectangle +(1,1);
  \draw[step=1cm,very thin,gray] (0,0) grid (4,4);
  \end{tikzpicture}
\hskip5pt
\begin{tikzpicture}[scale=0.2]
  \foreach \position in {(0,0),(3,3),(1,1),(3,0),(1,3),(0,3),(2,0),(3,1),(0,1)}
    \draw[fill=black] \position rectangle +(1,1);
  \draw[step=1cm,very thin,gray] (0,0) grid (4,4);
  \end{tikzpicture}
\hskip5pt
\begin{tikzpicture}[scale=0.2]
  \foreach \position in {(0,0),(1,0),(3,3),(0,3),(3,1),(0,1)}
    \draw[fill=black] \position rectangle +(1,1);
  \draw[step=1cm,very thin,gray] (0,0) grid (4,4);
  \end{tikzpicture}
\hskip5pt
\begin{tikzpicture}[scale=0.2]
  \foreach \position in {(0,0),(1,0),(3,0),(0,3),(0,1)}
    \draw[fill=black] \position rectangle +(1,1);
  \draw[step=1cm,very thin,gray] (0,0) grid (4,4);
  \end{tikzpicture}
\hskip5pt
\begin{tikzpicture}[scale=0.2]
  \foreach \position in {(0,0),(1,0),(2,3),(3,3),(0,3),(3,1),(2,1),(0,1)}
    \draw[fill=black] \position rectangle +(1,1);
  \draw[step=1cm,very thin,gray] (0,0) grid (4,4);
  \end{tikzpicture}
\hskip5pt
\begin{tikzpicture}[scale=0.2]
  \foreach \position in {(0,0),(1,0),(0,3),(2,0),(0,1)}
    \draw[fill=black] \position rectangle +(1,1);
  \draw[step=1cm,very thin,gray] (0,0) grid (4,4);
  \end{tikzpicture}
\hskip5pt
\begin{tikzpicture}[scale=0.2]
  \foreach \position in {(0,0),(1,0),(3,3),(1,1),(3,0),(1,3),(0,3),(3,1),(0,1)}
    \draw[fill=black] \position rectangle +(1,1);
  \draw[step=1cm,very thin,gray] (0,0) grid (4,4);
  \end{tikzpicture}
\hskip5pt
\begin{tikzpicture}[scale=0.2]
  \foreach \position in {(0,0),(1,0),(2,3),(3,3),(1,1),(3,0),(1,3),(0,3),(3,1),(2,1),(0,1)}
    \draw[fill=black] \position rectangle +(1,1);
  \draw[step=1cm,very thin,gray] (0,0) grid (4,4);
  \end{tikzpicture}
\hskip5pt
\begin{tikzpicture}[scale=0.2]
  \foreach \position in {(0,0),(1,0),(3,3),(1,1),(1,3),(0,3),(2,0),(3,1),(0,1)}
    \draw[fill=black] \position rectangle +(1,1);
  \draw[step=1cm,very thin,gray] (0,0) grid (4,4);
  \end{tikzpicture}
\hskip5pt
\begin{tikzpicture}[scale=0.2]
  \foreach \position in {(0,0),(1,0),(2,3),(3,3),(1,1),(3,0),(1,3),(0,3),(2,0),(3,1),(2,1),(0,1)}
    \draw[fill=black] \position rectangle +(1,1);
  \draw[step=1cm,very thin,gray] (0,0) grid (4,4);
  \end{tikzpicture}
\hskip5pt
\begin{tikzpicture}[scale=0.2]
  \foreach \position in {(0,0),(1,0),(2,3),(3,3),(1,1),(3,0),(1,3),(0,3),(3,1),(2,1),(1,2),(3,2),(0,1)}
    \draw[fill=black] \position rectangle +(1,1);
  \draw[step=1cm,very thin,gray] (0,0) grid (4,4);
  \end{tikzpicture}
\hskip5pt
\begin{tikzpicture}[scale=0.2]
  \foreach \position in {(0,0),(3,0),(0,3),(0,2),(3,1),(0,1)}
    \draw[fill=black] \position rectangle +(1,1);
  \draw[step=1cm,very thin,gray] (0,0) grid (4,4);
  \end{tikzpicture}